**Measuring spatially-resolved potential drops at semiconductor hetero-interfaces using 4D-STEM**

Varun Shankar Chejarla, Shamail Ahmed, Jürgen Belz, Jonas Scheunert, Andreas Beyer, Kerstin Volz*

Department of Physics and Materials Science Center, Philipps-University Marburg, Hans-Meerwein Str. 6, 35032 Marburg, Germany

*corresponding author: kerstin.volz@physik.uni-marburg.de

**Abstract**

Characterizing long-range electric fields and built-in potentials in functional materials at nano- to micrometer scales is of supreme importance for optimizing devices. E.g., the functionality of semiconductor heterostructures or battery materials is determined by the electric fields established at interfaces which can also vary spatially. In this study, we propose momentum-resolved four-dimensional scanning transmission electron microscopy (4D-STEM) for the quantification of these potentials and show the optimization steps required to reach quantitative agreement with simulations for the GaAs / AlAs hetero-junction model system. Using STEM the differences in the mean inner potentials (ΔMIP) of two materials forming an interface and resulting dynamic diffraction effects have to be considered. We show that the measurement quality is significantly improved by precession, energy filtering and a non-zone-axis alignment of the specimen. Complementary simulations yielding a ΔMIP of 1.3 V confirm that the potential drop due to charge transfer at the intrinsic interface is about 0.1 V, in agreement with experimental and theoretical values found in literture. These results show the feasibility of accurately measuring built-in potentials across hetero-interfaces of real device structures and its promising application for more complex interfaces of other polycrystalline materials on the nanometer scale.

**Introduction**

Built-in electric fields are the basis of many devices, e.g. transistors, solar cells or sensors.[1,2] In batteries, fields at interfaces can also occur and strongly influence their performance.[3–5] Those fields can extend from nanometers up to micrometers and must be distinguished from the sub-nanometer-ranged Coulomb field caused by the atoms in a specimen. With magnitudes in the range of a few MV/cm, the former ones are orders of magnitude weaker than atomic electric fields. For many applications there is a strong need to detect and quantify the longer-range electric fields occurring in devices. These fields often vary spatially as can be seen, e.g., for modern transistors that have three-dimensional architectures with different doping levels – and hence depletion region widths – in various directions.[6,7] Moreover, the lateral dimensions of these doped regions are becoming increasingly small.[6] In batteries, for example, potentials that limit or induce electron or ion transport, drop over the so-called Debye-length, extending over nanometers to hundreds of nanometers.[4,8] Due to the polycrystalline nature of several battery materials and their compositional and structural changes occurring in the material upon electrochemical cycling, the spatially resolved quantification of the characteristic Debye lengths is extremely challenging. Yet, it is mandatory to understand them to further improve energy storage materials.[4,8] Consequently, techniques providing a high spatial resolution capable of quantifying potential drops and electric fields across interfaces are of great relevance. In particular, modern transmission electron microscopes (TEM) provide a suitable resolution as well as the needed flexibility to address these challenges on a relevant length scale.

Accordingly, various TEM-based methods have already been applied to detect electric fields. By now especially electron holography, in which the electrostatic potential is measured,[9–13] and differential phase contrast imaging (DPC),[14–16] where the shift of the diffraction pattern due to the Coulomb interaction of the impinging electrons with the sample is detected,[17] were used. More recently, four-dimensional scanning transmission electron microscopy (4D-STEM) was utilized to detect the shifts of the electron probe in terms of momentum-resolved STEM (MRSTEM).[18,19] Acquiring a full diffraction pattern at each scan point has several advantages since various detection schemes can be applied, e.g., the center-of-mass (COM) of the diffraction pattern's intensity[18] or advanced edge detection and template matching schemes can be employed to track the shift of the pattern.[20,21] Moreover, 4D-STEM allows keeping track of spurious contributions to the COM signal, e.g., due to sample mistilt or strain. A technical review of the different methods was recently given by Addiego et al.[22]

Lately, we have shown that for prototypic GaAs-based p-n junctions, the key properties like built-in potential, field strength, doping levels or depletion width can be measured quantitatively by MRSTEM.[19] Similar results were also derived for other well-defined materials, i.e., Si-based p-n junctions, applying different imaging conditions.[23]

However, modern devices usually consist of more than one material and involve internal interfaces. At such hetero-interfaces, the mean inner potential (MIP), which is the volume average of the atomic electrostatic potentials,[24–26], also changes across the interface. Therefore, an impinging electron beam feels a potential difference of ∆MIP accross the width of the interface

Δx, corresponding to an apparent electric field. Consequently, a shift in the diffraction pattern is observed. This shift is also present at ideal non-charged interfaces and is caused only by the presence of different materials. Hence, quantitative electric field measurements at a hetero-interface, where additional built-in electric fields due to (un)intentional doping or charge redistributions can be present, are impossible without knowledge of the local MIP changes.

Accordingly, there have been several attempts to determine the MIP of a material experimentally, e.g., by electron holography.[13,27] However, especially in the case of electric fields at hetero-interfaces, dynamical effects and inhomogeneities from local thickness variations are challenging to handle.[13] At hetero-interfaces, the optimum phase reconstruction methods may require the combination of in-line and off-axis holography[13], making the evaluation of the data relatively complicated compared to momentum-resolved 4D-STEM data.

When using 4D-STEM to determine the effects of charges at interfaces in addition to the ΔMIP, dynamic diffraction also significantly complicates the interpretation of the data.[28–30] Strategies to reduce the impact of dynamic diffraction have already been proposed theoretically, e.g., by applying precession electron diffraction (PED).[29,30] However, quantitative values for the ΔMIP by 4D-STEM are still missing, making the determination of built-in potentials across heterojunctions using this technique impossible up to now.

In this work, we utilize advanced TEM techniques, such as a combination of 4D-STEM, energy filtering (EF) and PED in tandem with image simulation to quantitatively derive ΔMIP as well as the potential drop arising due to slight differences in the positions of the Fermi level across an interface. This is a prerequisite to address electric fields building up at interfaces due to charge redistributions. We systematically vary key experimental parameters like probe semi-convergence angle and precession angle and present optimum conditions, under which the potential drop at a hetero-interface can be measured quantitatively using 4D-STEM. We hence show that 4D-STEM can be applied – under carefully tuned experimental conditions – to derive electric fields and potential drops also at hetero-interfaces.

## Results

*Setup of the experiment and potential landscape across a heterointerface*

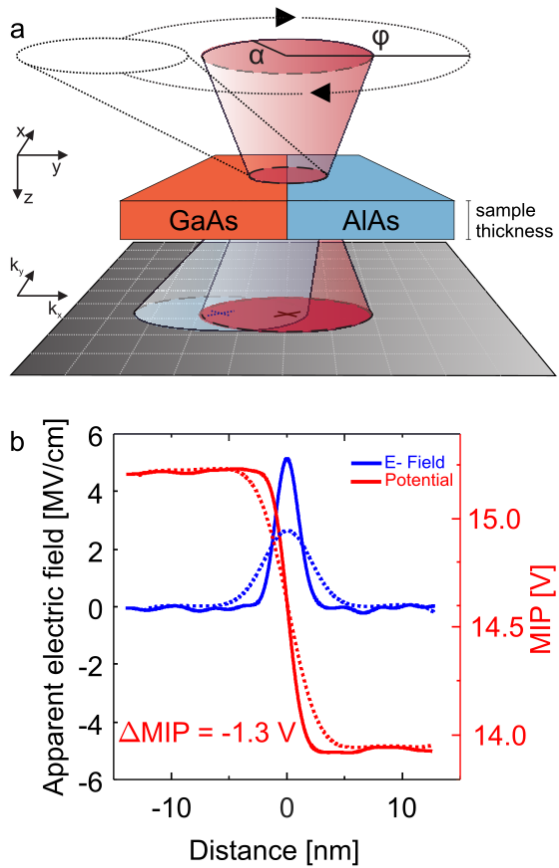

Figure 1: a) Schematic of the MRSTEM setup for heterointerfaces with a nearly parallel beam illumination. b) depicts the expected profiles of the apparent electric field (left y-axis, black line) and MIP (right y-axis, red line) across a GaAs / AlAs interface (if no built-in potential due to dipoles or extra charges at the interface is present). The solid and dashed lines denote convergence angles of 0.90 and 0.35 mrad, respectively.

The experimental MRSTEM setup used is sketched in **Figure 1a**. An electron beam with a convergence semi-angle of $\alpha$ is scanned over a TEM sample of thickness $t$. At each scan point a diffraction pattern is recorded so that any tilt of the beam introduced by electric fields can be tracked as a shift of the diffraction pattern. For example, the COM of the intensity of the unshifted and a shifted beam are marked by solid and dashed crosses, respectively. In addition, the electron beam can be precessed before hitting the sample using a precession angle $\varphi$, which is deprecessed below, leading to diffraction patterns with a reduced impact of dynamic diffraction.[26,31] Furthermore, by utilizing a energy filter we can improve the signal-to-noise ratio of the diffraction patterns by so-called zero-loss filtering.[32]

To prove the applicability of the method, we use a model hetero-interface, namely the abrupt, lattice-matched GaAs / AlAs (001) interface. The built-in potential landscape at this interface will be discussed later.

The MIPs, being the volume average of the electrostatic potentials for electrons in a solid, have larger absolute values for materials with a larger number of protons, i.e., in our case GaAs.[33] Hence, at the same interface, the MIP landscape exhibits a step with a height of $\Delta V$ = ΔMIP. Scanning across this interface, the electron probe will be attracted by the GaAs and a negative COM shift is a result, as illustrated in Figure 1a. As expected, this will result in a positive apparent electric field and a potential change across the interface. Consequently, using MRSTEM, we can quantify the sign and the value of the potential drop ΔMIP across an interface. This potential distribution is plotted for a perfect GaAs / AlAs interface without any dipole at the interface, as shown by the red line in Figure 1b. This curve was derived by averaging the slice potentials containing the isolated atomic potentials[34] from a multi-slice STEM simulation[35] over one crystal unit cell.[29] The value of the potential drop $\Delta V$ = ΔMIP determined this way equals -1.30 V. This number is in reasonable agreement with the density functional theory (DFT) calculations reported in[33,36] as well as holographic data.[37] The apparent electric field $E = \Delta V/\Delta x$ is plotted as a blue line in Figure 1b. The width $\Delta x$ of this transition depends on the actual width of the interface as well as on the width of the impinging electron probe and its broadening within the sample. Consequently, the magnitude of the apparent field also depends on the imaging conditions used in the MRSTEM measurements. This is highlighted by the field profile derived assuming a broader electron probe, which is shown as a dashed blue line in Figure 1b. A broader probe can be realized experimentally by reducing its semi-convergence angle, e.g., by choosing a smaller condenser aperture. However, the height of the potential step derived by integrating over the electric field is unaffected by the probe size. Experimental data for two different semi-convergence angles further proving this point is shown in **Figure S1**. Hence, the potential is the physically meaningful quantity to focus on in the following. It should be pointed out again that this "apparent" electric field is also present at ideal charge-neutral interfaces solely caused by the difference in MIP of the two materials. However, in our experiments, the absolute MIP values of both materials are not quantifiable due to a lack of reference. Hence, we chose to integrate the apparent electric field at the interface for obtaining the potential depicted as red curves in the experimental plots.

In the following, we will show how to reliably measure this potential step across the GaAs / AlAs interface and discuss our results. It should be emphasized that - if an additional field is present due to electric dipoles or charges at the interface - the shift of the COM induced by this field will be overlaid over the COM shift induced by the ΔMIP. Hence, also this additional field can be quantified when knowing ΔMIP.

*Experimental Results*

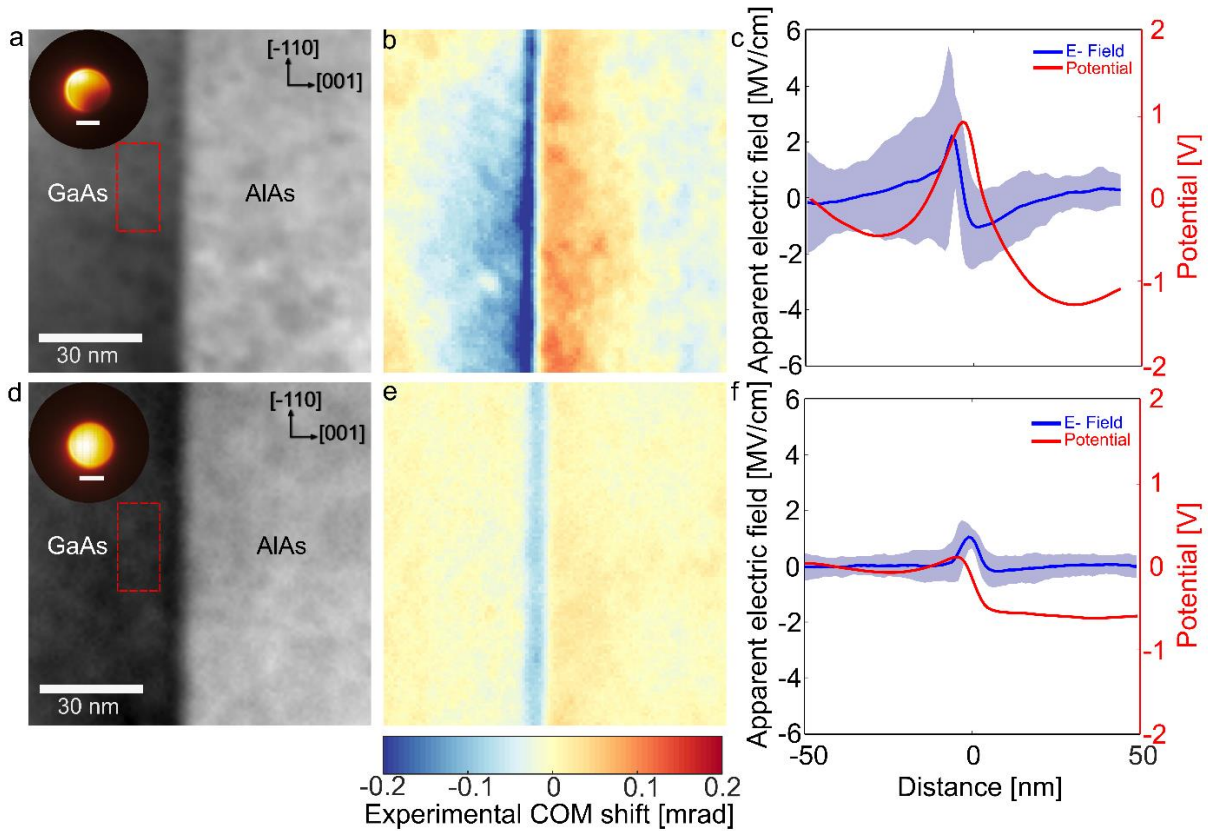

Figure 2: COM shift and apparent electric field investigations without (a-c) and with precession (d-e; precession angle 0.4°) for a scanning parallel beam having 0.90 mrad semi-convergence angle in [-110] zone axis: a) and d) show VBF images acquired from the 4D datasets across the GaAs / AlAs interface. The insets in both figures are the PACBEDs acquired from the regions indicated with the dashed red rectangles. The scale in the insets corresponds to 1 mrad. b) and e) are the [001] COM component maps in false color-coding acquired without precession b) and with precession e), respectively. c) and f) are the apparent electric field profiles in MV/cm (left y-axis) and the potential in V (right y-axis). The shaded regions in the profiles indicate the standard deviations of the electric field over the FOV of the data in b) & e).

**Figure 2** shows the experimental COM shift and its apparent electric field across the GaAs / AlAs hetero-interface in [-110] zone axis projection. An electron probe with a semi-convergence angle of 0.90 mrad is scanned across the interface unprecessed (a – c) and precessed with an angle of 0.4° (d – f), respectively. Figures 2a and 2d show the corresponding STEM Virtual Bright Field (VBF) images of the hetero-interface. A dip in intensity due to dynamical diffraction can be observed at the bottom right of the Position-Averaged Convergent Beam Electron Diffraction (PACBED) of the unprecessed dataset (a). These dynamical diffraction features can change their position in the bright field (BF) disc drastically if there are only the slightest variations in sample thickness and local misorientation (e.g., due to very small sample bending) over the field-of-view (FOV) of a TEM sample.[31] Such local sample thickness and misorientation variations are extremely hard to avoid. This implies that the intensity redistribution in the BF disc cannot be solely related to the apparent electric field or potential at the interface if dynamical diffraction conditions vary over the FOV. Figure 2b shows the [001] COM component obtained by scanning the nearly parallel probe across the hetero-interface. The impact of the dynamic effects described is evident in this image. As described in the experimental section, an apparent electric field is calculated using the experimentally determined COM shift and displayed as a blue line in Figure 2c. Integrating the apparent electric field profiles leads to a potential drop at the interface which is shown in the right y-axis of Figure 2c. The COM shift measured was set to zero in a reference region, i.e., the GaAs layer on the left-hand side. Accordingly, the potential curve starts at 0 V in GaAs and not at its theoretical value of 15.2 V, as shown in Figure 1b. The shaded areas surrounding the field graphs give an error estimation. They represent the standard deviation of all pixels values along a line parallel to the interface in (b). The corresponding error values for the potentials are derived by adding the electric field data of 20 pixels along the interface and determining the standard deviation of the potentials derived from these fields along the interfaces. Since the standard deviation of the potential is a suitable figure of merit for the accuracy of the method, the values are also compiled in Table 1 at the end of the experimental section for all experimental conditions studied. The data shown in Figure 2 reflects the dynamic effects manifesting as intensity variations in the BF disc. By precessing the beam with an angle of 0.4° around the optic axis, the effects of dynamic diffraction are reduced, as concluded from the rather homogeneous intensity distribution in the PACBED in the inset of Figure 2d. The optimization of the precession angle is shown in the supplementary **Figure S2**. It can be seen that low precession angles (0.2° and 0.3°) show very pronounced dynamic diffraction effects, adversely affecting the COM shift evaluation. Hence, higher precession angles are prefered to cancel these dynamic effects. However, very high precession angles lead to precession-induced two-fold astigmatism that reduces the spatial resolution.[38] For that reason, 0.4° is chosen to be the optimum precession angle. Figure 2e depicts the [001] COM component obtained by scanning the precessed beam across the hetero-interface. Figure 2f shows the apparent electric field line profile across the interface from the COM component and the calculated integrated profile (potential on the right y-axis). Figure 2f proves that precessing the probe can significantly reduce the dynamic diffraction effects. This is mainly reflected by the standard deviation for the values of the potential at the interface which is reduced from 0.74 V to 0.13 V. The evaluation of the potential drop, however, is still conflicted by the strong dynamic diffraction effects at the interface, extending also into the respective materials.

Thermal diffuse and inelastic scattering have an important contribution to TEM measurements, especially when the TEM lamellae are thick. This also impacts the evaluation of the COM shift from 4D datasets. Inelastic events at scattering angles close to the convergence angle, e.g., from plasmon excitations,[39–41] on the direct electron detector show a statistical distribution that influences the COM shift especially for low electron doses. To make the COM investigations more robust, EF with a precessing beam (0.4°) is investigated in the next step of optimization. This minimizes the impact of inelastic scattering events and increases the signal-to-noise ratio in the 4D datasets. For this purpose, the so-called zero-loss peak, which contains electrons that encountered only small energy losses, is selected by inserting a slit aperture into the spectrum plane. Figure 3 shows the scanning-precessed (nearly) parallel beam data with zero-loss EF in [-110] zone axis at 0.90 mrad semi-convergence angle. Figure 3a shows the VBF image derived from that data. Compared to Figure 2e, the measured EF COM shift in Figure 3b shows less scatter which is especially visible in the regions away from the hetero-interface. The calculated apparent electric field profile from the measured COM shift and the calculated potential from the apparent electric field is shown in Figure 3c. Significant improvement in the COM shift is observed due to the EF with precession at the hetero-interface. The measured value for potential without EF is considerably lower than expected at the interface. With EF, however, the value for the potential has increased to 1.26 V. Additionally, the standard deviation for the values of potentials is reduced from 0.13 V (without EF (from Figure 2f)) to 0.11 V (with EF (from Figure 3c)) as shown in Table 1.

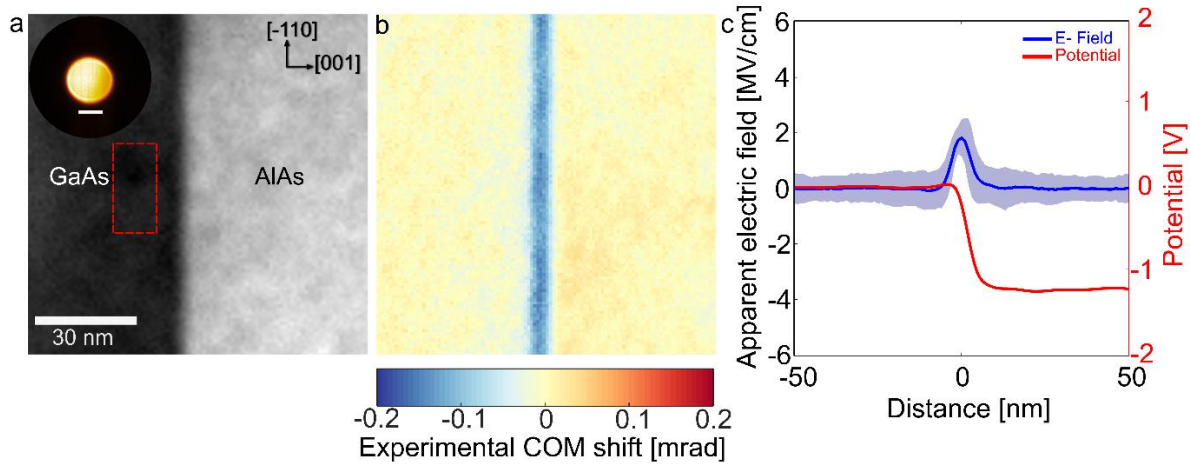

Figure 3: Impact of EF for 0.90 mrad semi-convergence angle with precession (angle 0.4°): a) shows a VBF image acquired from the 4D dataset across the GaAs / AlAs interface. The inset in the figure depicts the PACBED acquired from the region indicated with the dashed red rectangle. The scale in the inset corresponds to 1 mrad. b) shows the [001] COM component map in false color-coding. c) shows the experimental apparent electric field profile in MV/cm (left y-axis) and the potential in V (right y-axis). The shaded region in the profile indicates the standard deviations of the electric field over the FOV of the data in b).

Even though precession suppresses the effect of dynamic diffraction to a large extent and EF helps in suppressing statistical effects of inelastic scattering, dynamical diffraction effects can still be observed in the PACBEDs in zone-axis orientation (e.g., PACBEDs shown in Figures 2d and 3a). To eliminate the diffraction effects from the crystalline sample in the best possible manner, we tilt the sample out-of-zone axis condition.[23,42] The sample is tilted by around 7° parallel to the hetero-interface without projecting the interface.

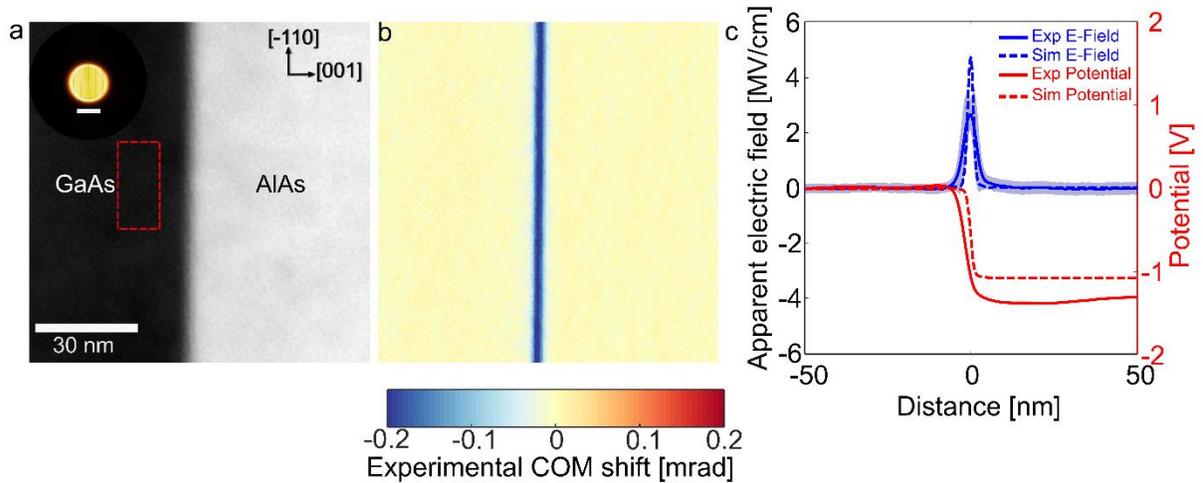

Figure 4: Impact of out-of-zone axis condition for scanning precessed EF 4D dataset with 0.90 mrad semi-convergence angle: a) shows a VBF image acquired from the 4D dataset across the GaAs / AlAs interface. The inset in the figure shows the PACBED acquired from the region indicated with the dashed red rectangle. The scale in the inset corresponds to 1 mrad. b) shows the [001] COM component map in false color-coding. c) shows the experimental (solid lines) and simulated (dashed lines) apparent electric field profiles (left y-axis) and the potential profiles (right y-axis). The shaded regions in the profile indicate the standard deviations of the experimental electric field over the FOV of the data in b). The simulations assume only the effects of ΔMIP.

**Figure 4** shows the EF precessed (0.4°) scanning data acquired in the out-of-zone axis condition for a semi-convergence angle of 0.90 mrad. The VBF and COM shift data are shown in Figures 4a and 4b, respectively. It can be seen that the dynamic diffraction effects due to zone axis conditions present in the previous data is significantly reduced when the sample is tilted. Accordingly, the COM data shows fewer fluctuations over the field of view. This is reflected in the experimental apparent electric field profile calculated from the COM shift that is shown as a blue line in Figure 4c. In addition, Figure 4c also depicts the profile of the potential on the right y-axis . The standard deviation of the potential is significantly reduced to 0.02 V compared to the ones derived from Figures 2c and 2f as well as Figure 3c. This is highlighted in Table 1, where the standard deviations derived from the different experiments are collected. Moreover, the electric field profile of the off-axis data (Figure 4c) is much sharper compared to the previous data. This is most likely caused by a reduced beam broadening in the off-zone-axis condition. The corresponding profiles derived from the simulations, assuming only ΔMIP and no built-in potential at the heterointerface, are also shown in Figure 4c as dashed lines. The simulated profile appears significantly sharper due to the higher spatial resolution in the simulation, as described in the experimental section. The impact of varying the convergence angle in the out-of-zone axis condition is shown in Figure S1. Using a lower semi-convergence angle of only 0.35 mrad, the COM shift and apparent electric field profiles are broader due to the larger beam diameter. However, the actual physical quantity, i.e., the potential of 1.38 V ± 0.02 V, is in quantitative agreement with the values at a higher convergence angle (Figure 4).

**Table 1:** Potential drops in Volts across a GaAs / AlAs interface calculated from experimental COM shifts by using an electron probe with a semi-convergence angle of 0.90 mrad in different probing conditions. Errors are derived from standard deviation.

|  | Without precession in zone axis | With a precession angle of 0.4° in zone axis | With EF and a precession angle of 0.4° in zone axis | Tilted out-of-zone axis with EF and with precession angle 0.4° |
|---|---|---|---|---|
| Potential drop [V] | 1.92 ± 0.74 | 0.7 ± 0.13 | 1.26 ± 0.11 | 1.41 ± 0.02 |

*Discussion*

In summary, we have shown that by carefully conducting a MRSTEM experiment the potential drop across a GaAs / AlAs interface can be reproducibly determined to be 1.41 V with a very low error and standard deviation of the data of only ± 0.02 V. The theoretical value for only the ΔMIP at this interface has been calculated to be 1.30 V.

We will show in the following that this small deviation can be explained by the built-in potential at the GaAs / AlAs interface. If grown without doping, this interface is certainly amongst the ones with the lowest built-in potentials. However, even at this interface, dipoles are present due to the different electronegativity of the two materials leading to a charge transfer across the interface which results in a small electric field and the potential drop across the interface. Hence, a shift of the COM of the scanned electron beam is observed across the interface. Experimentally measured valence band offsets (VBO) between GaAs and AlAs are, on average, 0.53 eV.[43–45] Different theoretical approaches yield numbers between 0.36 - 0.54 eV for this quantity.[46–50] Together with the band gaps being 1.52 eV for GaAs (at the Γ-point) and 2.24 eV for AlAs (at the X-point), a built-in potential between 0 - 0.18 V will develop even at undoped GaAs / AlAs interfaces.[43] As positive charges will accumulate at the GaAs side of the interface and negative ones at the AlAs side of the interface, the electron beam will also be attracted towards the GaAs at the interface, and a negative COM shift will be observed. This will result in an observed positive electric field and a potential drop across this interface. A schematic of this scenario is shown in **Figure S3**.

The COM shift from the MIP difference and from the built-in potential across a heterointerface will overlay each other in the MRSTEM data, resulting in a quantitative agreement for our measured value, having contributions from ΔMIP as well as the built-in potential.

It should be noted that MRSTEM does not only give quantitative values for ΔMIP and the built-in potential across an interface but also that the experimental error of this method (which we estimated by the maximum standard deviation) is significantly reduced compared to electron holography data. This makes the presented method very promising not only to determine the ΔMIP at hetero-interfaces but also to measure real electric fields at hetero-junctions.

**Conclusion**

Built-in potentials and the resulting electric fields at interfaces are the basis of many devices and their quantification at high spatial resolution is of great relevance. We show that the momentum transfer of such an electric field on a relativistic electron beam can be quantitatively evaluated to yield these fields and the corresponding potentials. Dynamic diffraction effects resulting from MIP changes across hetero-interfaces, which are the greatest hurdle in quantifying built-in potentials across internal interfaces, are shown to be suppressed by a combination of advanced STEM techniques, namely precessing the electron probe around the optical axis and zero-loss filtering the signal. Moreover, tilting the sample to out-of-zone-axis conditions significantly increases the robustness of the quantification of the potentials. Consequently, the mean inner potential difference and the small built-in potential forming across a model GaAs / AlAs interface have been quantitatively defined, underlining the tremendous potential of the MRSTEM technique.

**Experimental Section**

The GaAs / AlAs hetero structure was grown via metal-organic vapor phase epitaxy (MOVPE) in an AIX 200 rector. Growth conditions have been chosen to avoid any unintentional doping of the layers. An electron transparent sample was prepared via Focused-Ion Beam (FIB) preparation methods using a JEOL JIB-4601F FIB/Scanning Electron Microscope (SEM). At first, a thin layer of tungsten was deposited via an electron beam on the sample's surface. Then an about 4 µm thick tungsten protection layer was deposited using a 30 kV Ga-ion beam in order to protect the sample surface. The sample was attached to a TEM grid and thinned down initially to about 500 nm with a 30 kV Ga-ion beam. The accelerating voltage of the Ga-ions was reduced progressively to further thin down the sample. Final polishing was done using a 1 kV Ga-ion beam to reach a final TEM sample thickness of 245 ±10 nm, which was measured by Electron Energy Loss Spectroscopy (EELS). Before the TEM investigation, the sample was stored in an Ar-filled glovebox to prevent oxidation. The transfer between the glovebox and the TEM was optimized to minimize air exposure as much as possible.

The TEM measurements were carried out on a double Cs-corrected JEOL JEM-2200FS operated at 200 kV. Two different condenser lens apertures (CLAs) were used in this study, providing semi-convergence angles of about 0.90 mrad and 0.35 mrad, respectively. The semi-convergence angles were precisely measured for each CLA after every measurement session to correct the slight variations between different measurement sessions. A sketch of the experimental setup is shown in Figure 1a. In the microscope's TEM mode, the probe was first focused on the sample and then scanned using the external P2000 scanning unit of the NanoMegas PED system. Four-dimensional datasets, with dimensions $(x,y,k_x,k_y)$, were acquired using a pnCCD pixelated detector, with 1 ms dwell time in full-frame mode, giving frames of dimensions ($k_x$ = 264, $k_y$ = 264) in the momentum space. Precession was applied with a frequency of 1000 Hz using the P2000 scan unit. This results in one precession cycle per frame for the datasets with precession. A camera length of 100 cm was used, which resulted in an angular density of about 0.025 mrad/pixel in momentum space. For the datasets where energy filtering was applied, an energy slit of about 13 eV width was placed at the zero-loss region of the energy spectrum.

To increase the signal-to-noise ratio, 4D datasets of dimensions (x - 512, y - 512, $k_x$ - 264, $k_y$ - 264) were acquired and, four frames in the momentum space (corresponding to two adjacent pixels in the x direction and two adjacent pixels in the y direction) were averaged, yielding final datasets with dimensions (256, 256, 264, 264). A circular mask of around twice the radius of the direct beam disk's radius was applied to each frame to obtain the COM of the diffraction pattern to avoid contributions from other diffraction spots. A linear 2D background was fitted to the COM components for each dataset to correct for artifacts due to de-scan.[19] The apparent electric field E was calculated from the COM by E = $- h * \frac{sin(COM)}{\lambda} * \frac{v}{t*e}$, with h being Planck´s constant, λ being the wavelength, v the velocity and e the charge of the electron and t being the TEM sample thickness.[18] To obtain the apparent electric field profiles across an interface, the COM data were averaged in the perpendicular direction along the interface. The potential was then derived by integrating the apparent electric field across the interface, choosing 0 V as a reference point.

Complementary simulated 4D datasets were generated using the multi-slice algorithm implemented in the STEMsalabim code,[35] assuming the same imaging conditions used in the experiments. As input, [110] oriented GaAs super cells with dimensions of 34 x 34 x 300 nm³ were generated. A defined change in MIP was realized by replacing the Ga atoms in the right half of the cell with Al, keeping the lattice constant and Debye-Waller factors of GaAs. To model the impact of PED, 24 individual specimen tilts were simulated and averaged before evaluating the simulated COM data analogously to the experimental ones. It is worth noting that in the experiment, the beam is tilted and not the sample, which leads to a reduction of the spatial resolution of the experiment in comparison to the simulation due to the imperfection of the TEM's probe-forming lens system.

**Acknowledgement:**

Funding by the German Research Foundation (DFG) in the framework of SFB 1083 "Structure and Dynamics of Internal Interfaces" (project number 223848855) as well as the Bundesministerium für Bildung und Forschung (BMBF) in the framework of "H2Demo" (project number03SF0619F) is gratefully acknowledged.

**Supplementary Figures:**

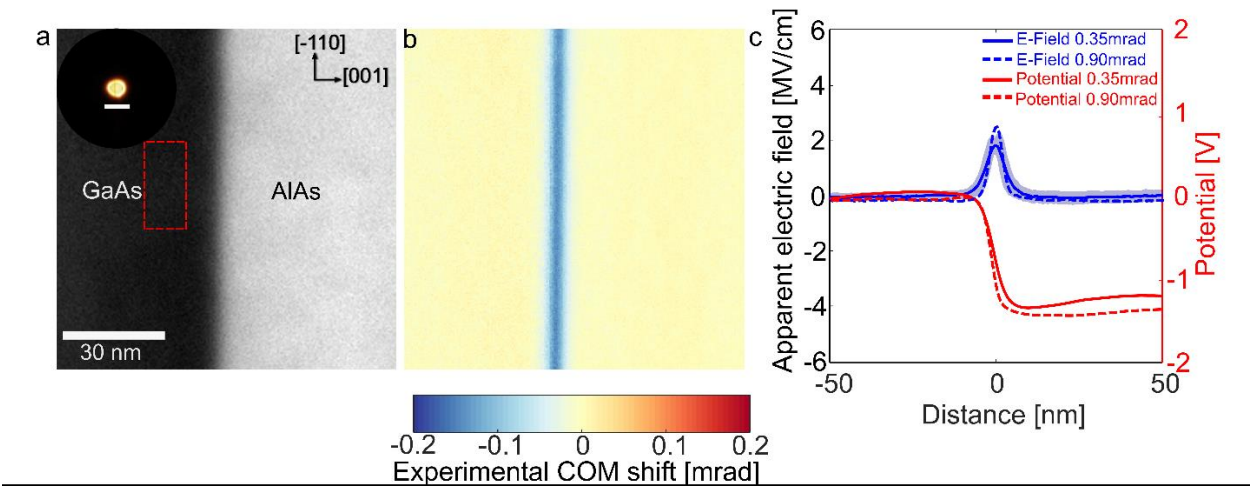

Figure S1: Impact of convergence angle in out-of-zone axis condition for scanning precessed EF 4D-data set with 0.35 mrad semi-convergence angle: a) shows a VBF image acquired from the 4D dataset across the GaAs / AlAs interface. The inset in the figure shows the PACBED acquired from the region indicated with the dashed red rectangle. The scale in the inset corresponds to 1 mrad. b) shows the [001] COM component map in false color-coding. c) shows the apparent electric field profiles for 0.35 mrad and 0.90 mrad semi-convergence angles in MV/cm (left y-axis) and the potential profiles in V (right y-axis). The shaded region around the solid blue line profile indicates the standard deviations of the electric field over the FOV of the data in b) for 0.35 mrad semi-convergence angle. The standard deviations of the electric field for 0.90 mrad semi-convergence angle (dashed blue line) is omitted for better visualization, as it is already shown in Figure 4c.

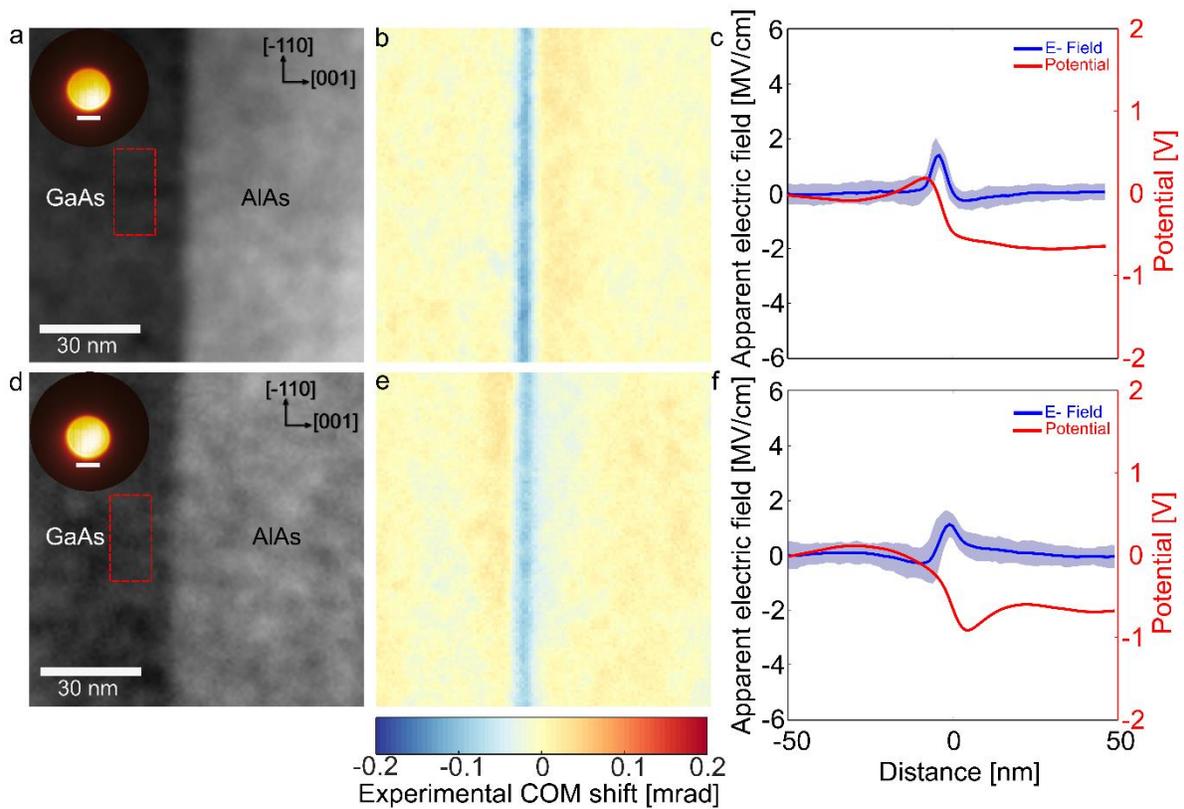

Figure S2: COM shift and apparent electric field investigations with precession angle 0.2° (a-c) and with precession angle 0.3° (d-f) for a scanning beam having 0.90 mrad semi-convergence angle in [-110] zone axis: a) and d) show VBF images acquired from the 4D datasets across the GaAs / AlAs interface. The insets in both figures are the PACBEDs acquired from the regions indicated with the dashed red rectangle. The scale in the insets corresponds to 1 mrad. b) and e) are the [001] COM component maps in false color-coding. c) and f) are the apparent electric field profiles (left y-axis) and the potential profiles in V (right y-axis). The shaded regions in the profiles indicate the standard deviations of the electric field over the FOV of the data in b) & e). As observed in the COM maps and the profiles for both 0.2° and 0.3° precession angles, the dynamic effects away from the interface have a significant impact on the apparent electric field and on the potential.

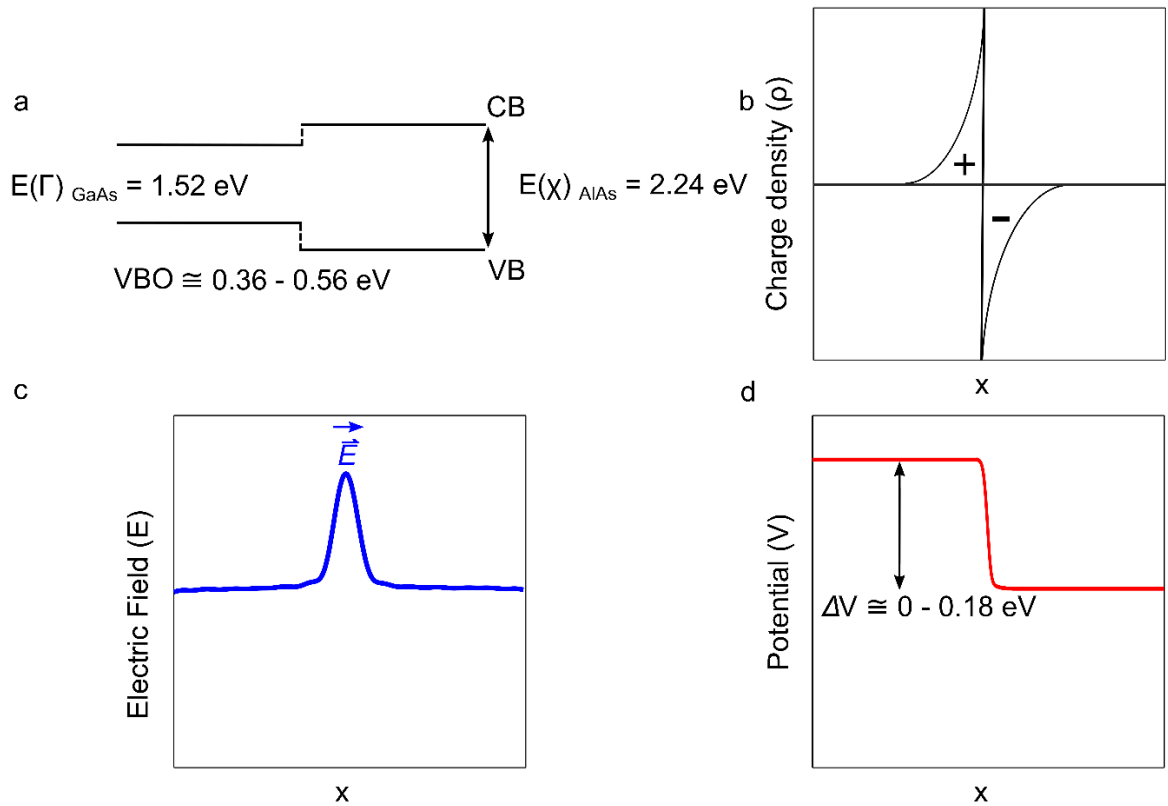

Figure S3: a) Band alignment at the GaAs / AlAs interface. The relevant numbers are VBO 0.36 - 0.54 eV, bandgaps: GaAs 1.52 eV, AlAs 2.24 eV b) for VBO > 0.36 eV positive charge is transferred to GaAs, negative to AlAs, this results in an electric field, depicted in c). The resulting built-in potential shown in d) is between 0 - 0.18 eV, depending on VBO.